\documentstyle[amsfonts,amsmath,twoside,multicols,epsf,times]{article}
\input amssym.def
\input amssym
\begin{document}

\def\nl{\noindent}
\def\bea{\begin{eqnarray}}
\def\nonu{\nonumber }
\def\eea{\end{eqnarray}}
\def\psla{\rlap \slash}
\newcommand{\be}{\begin{equation}}
\newcommand{\ee}{\end{equation}}
\def\sla#1{\rlap\slash #1}
\def\dk{{d^2 k_{\perp} dk^{+} \over (2 \pi)^3 2 k^{+}}}
\def\dx{{1 \over 2} \int dx^{-} d^{2} x^{\perp}}
\newcommand{\bu}{$\bullet$}

\pagestyle{myheadings}
\markboth{Brazilian Journal of Physics, vol.  34, no. 3, September, 2004}
{J. P. B. C. de Melo and T. Frederico }
\title{Spin-1 Particle in the Light-Front Approach}
\author{J. P. B. C. de Melo$^a$ and T. Frederico$^b$\\
{\small\it  $^a$Instituto de F\'\i sica Te\'{o}rica,
Universidade Estadual
Paulista, 01405-900, S\~{a}o Paulo, SP, Brazil} \\
{\small\it $^b$Departamento de F\'\i sica, Instituto Tecnol\'ogico de
Aeron\'autica, Centro T\'ecnico Aeroespacial,}\\
{\small\it 12.228-900, S\~ao
Jos\'e dos Campos, S\~ao Paulo, Brazil}  }
\date{Received on October 07, 2003.}
\maketitle
\begin{abstract}
The electromagnetic current of spin-1 composite particles does not
transform properly under rotations if only the valence
contribution is considered in the light-front model. In
particular, the plus component of the current, evaluated only for
the valence component of the wave function, in the Drell-Yan frame
violates rotational symmetry, which does not allow a unique
calculation of the electromagnetic form-factors. The prescription
suggested by  Grach and Kondratyuk [Sov. J. Nucl. Phys. 38, 198
(1984)] to extract the form factors from the plus component of the
current, eliminates contributions from pair diagrams or zero
modes, which if not evaluated properly cause the violation of the
rotational symmetry. We address this problem in an analytical and
covariant model of a spin-1 composite particle.
\end{abstract}

\begin{multicols}{2}
\baselineskip=11.5pt

\section{Introduction}

Light-front models are useful to describe hadronic bound states,
like mesons or baryons due to its particular boost
properties~\cite{Terentev76,Brodsky98}. However, the light-front
description in a truncated Fock-space breaks the rotational
symmetry because the associated transformation is a dynamical
boost~\cite{Pacheco97,Pacheco98,Pacheco992}. It is a formidable
task to study the properties under dynamical boosts in light-front
quantization~\cite{Brodsky98}. Therefore, an analysis with
covariant analytical models, can be useful to pin down  the main
missing features in a truncated light-front Fock-space description
of the composite system. In this respect, the rotational symmetry
breaking of the plus component of the electromagnetic current, in
the Drell-Yan frame, was recently studied within an analytical
model for the spin-1 vertex of a composite two-fermion bound
state~\cite{Pacheco97,Pacheco98}. It was shown that, if pair term
contributions are ignored in the evaluation of the matrix elements
of the electromagnetic current, the covariance of the form factors
is lost~\cite{Pacheco97,Pacheco98,Pacheco992,JI2001}. The complete
restoration of covariance in the form factor calculation is found
only when pair terms or zero modes contributions to the matrix
elements of the current are
considered~\cite{Pacheco98,Pacheco992,Naus98,Pacheco99}.

The extraction of the electromagnetic form-factors of a spin-1
composite particle from the microscopic matrix elements of the
plus component of the current ($J^+=J^0+J^3$) in the Drell-Yan
frame (momentum transfer $q^+=q^0+q^3=0$), based only on the
valence component of the wave function, presents ambiguities due
to the lacking of the rotational invariance of the current
model~\cite{Inna84,Inna89}. In the Breit-frame with  momentum
transfers along the transverse direction (the Drell-Yan condition
is satisfied) the current $J^+$ has four independent matrix
elements, although only three form factors exist. Therefore, the
matrix elements satisfies an identity, known as the angular
condition~\cite{Inna84}, which is violated.

Several extraction schemes for evaluating the form-factors were
proposed, and in particular we consider the suggestion made in
Ref.~\cite{Inna84}. It was found in a numerical calculation  of
the $\rho$-meson electromagnetic form factors considering only the
valence contribution~\cite{Pacheco97}, that the prescription
proposed by~\cite{Inna84} to evaluate the form-factors, produced
results in agreement with the covariant calculations.
In Ref.~\cite{Pacheco97}, it was used an analytical form of the
$\rho$-quark-antiquark vertex.  Later, in Ref.~\cite{JI2001}, it
was shown that the above prescription eliminates the pair diagram
contributions to the form factors, using a simplified form of the
model, when the matrix elements of the current were evaluated for
spin-1 light-cone polarization states. This nice result was
thought to be due to the use of the particular light-cone
polarization states. Here, we will show in a straightforward and
analytic manner that the cancellation of the pair contribution in
the evaluation of the form factors using the prescription from
Ref.~\cite{Inna84} also happens for the instant form polarization
states in the cartesian representation, generalizing the previous
conclusion~\cite{JI2001}. Our aim, is to expose in a simple and
detailed form, how the pair terms appear in the matrix elements of
the current evaluated between instant form polarization states,
and their cancellation in the form factors using the correct
prescription. Therefore, we conclude that this property is more
general than realized before.

We should observe that, in the case of spin-0 composite particles
(like the pion) with the correspondent form of the analytical
model, the plus component of the electromagnetic current in the
Breit-frame, with $q^+=0$, does not have contributions from pair
terms~\cite{Pacheco99,Pacheco2002}. It is enough to evaluate the
valence part of the matrix element of the current to reproduce the
covariant result. We remind that in the case of other components
of the current, like $J^-=J^0-J^3$, the pair term contributes to
the matrix element of the current \cite{Pacheco99}.

This work is organized as  follows. In Sect. II, we present the
light-front model of the spin-1 particle, and evaluate the matrix
elements of the plus component of the electromagnetic current in
the Breit-frame with $q^+=0$. We separate out the pair terms in
the matrix elements using the pole dislocation
method~\cite{Pacheco98,Naus98,Pacheco992,Pacheco99}. In  Sect.
III, we show the cancellation of the pairs terms when the form
factors are evaluated with the instant form polarization states
with the prescription suggested by Grach and
Kondratyuk~\cite{Inna84}. In Sect. IV, we present our conclusion.

\section{Light-front model  }

The electromagnetic (e.m.) current has the following general form
for spin-1 particles \cite{Frankfurt79}:
\begin{eqnarray}
&&J_{\alpha \beta}^{\mu}=[F_1(q^2)g_{\alpha \beta} -F_2(q^2)
\frac{q_{\alpha}q_{\beta}}{2 m_\rho^2}] (p^\mu + p^{\prime \mu})
\nonumber \\&& - F_3(q^2) (q_\alpha g_\beta^\mu- q_\beta
g_\alpha^\mu) \ , \label{eq:curr1}
\end{eqnarray}
where $m_v$ is the mass of the vector particle, $q^\mu$ is the
momentum transfer, $p^\mu$ and $p^{\prime  \mu}$ is on-shell
initial and final momentum respectively. From the covariant form
factors $F_1$, $F_2$ and $F_3$, one can obtain the charge ($G_0$),
magnetic ($G_1$) and quadrupole ($G_2$) form factors (see
e.g.~\cite{Pacheco97}).

In the impulse approximation for the elastic photo-absorption
amplitude (represented by a Feynman triangle diagram), the matrix
elements of the e.m. current, $J^{+}$ is written as~\cite{Pacheco97}:
\begin{eqnarray}
&&J^+_{ji}=\imath  \int\frac{d^4k}{(2\pi)^4}
 \frac{Tr[  \ ]_{ji}~\Lambda(k,p^\prime)\Lambda(k,p) } {((k-p)^2
- m^2+\imath\epsilon) (k^2 - m^2+\imath \epsilon)} \nonumber \\
&&\times \frac{1}{((k-p^\prime)^2 - m^2+\imath \epsilon)}   \ ,
\label{jcurrent}
\end{eqnarray}
where the trace is given by:
\begin{eqnarray}
&&Tr[  \ ]_{ji}=Tr[\epsilon^{\prime\alpha}_j
\Gamma_{\alpha}(k,k-p^\prime) ({\sla k}-\sla{p}^\prime +m)
\gamma^{+} \nonumber \\
&&\times \ (\sla{k}-\sla{p}+m) \epsilon^\beta_i
\Gamma_{\beta}(k,k-p) (\sla{k}+m)] \ , \label{trace}
\end{eqnarray}
with $\gamma^+=\gamma^0+\gamma^3$. The polarization four-vectors
of the initial and final states are $\epsilon_i$ and
$\epsilon^\prime_j$, respectively. The regularization function
$\Lambda(k,p)=N/((p-k)^2-m_R^2+\imath \epsilon)^2$ is used to keep
finite the photo-absorption amplitude. The regularization
parameter is $m_R$.

In this work, we make use of a simple Dirac structure in the
spinor space for the vertex $\Gamma_{\alpha}(k,k-p)$, which is
written as $\Gamma$ = $\Gamma^{\prime}$=$\gamma^{\mu}$. Although,
this vertex has a simple structure, it originates $S$ and $D$
states in the relative motion of the fermions which composes the
spin-1 particle. The full form of the vertex, applied previously
to study the $\rho$-meson e.m. form factors, has one more term
which puts the quarks on the relative $S$ state \cite{Pacheco97}.
With only the $\gamma^\mu$ structure, the trace of Eq.
(\ref{trace}) becomes:
\begin{eqnarray}
Tr[ \ ]_{ji}=Tr[     \sla{\epsilon^{\alpha}_f}
(\sla{k}-\sla{p^\prime} +m) \gamma^{+} (\sla{k}-\sla{p}+m)
\sla{\epsilon^{\alpha}_i} (\sla{k}+m)]\ . \label{trace+}
\end{eqnarray}
Using the  light-front coordinates as $k^+=k^0+k^+$, $k^-=k^0-k^3$
and $k_{\perp}=(k_x,k_y)$), the  $k^-$ dependence in the trace is
separated. Following ~\cite{Pacheco992}, we treat the $k^-$
factors which come from the instantaneous  term of the fermion
propagators carefully, because they can cause the violation of the
rotational symmetry of the plus component of the e.m. current.
These contributions are related to the presence of pair terms in
the matrix elements of the e.m. current in the limit where
$q^+=\delta^+$ goes to zero~\cite{Pacheco992}.

Writing only the $k^-$ dependence of the trace, one has:
 \be Tr[ \ ]^{Bad}_{ji} = \frac{k^-}{2} \
 Tr[\sla{\epsilon}^{\prime\alpha}_j
(\sla{k}-\sla{p^\prime} +m) \gamma^{+} (\sla{k}-\sla{p}+m)
\sla{\epsilon}^{\alpha}_i \gamma^+ ] \  . \ee The terminology
"Bad" is used here to indicate the possible contribution of  pair
terms.

The matrix elements of the e.m. current are calculated in the
Breit frame with the momentum transfer $q^\mu=(0,q_x,0,0)$,
$p^\mu=(p^0,-q_x/2,0,0)$ for the spin-1 particle initial state and
$p^{\prime\mu}=(p^0,q_x/2,0,0)$ for the final state. Introducing
the useful definition $\eta=q^2/4 m_{v}$, we have
$p^0=m_{v}\sqrt{1+\eta}$. The instant-form polarization
four-vectors in the cartesian representation are given by
$\epsilon^\mu_x=(-\sqrt{\eta},\sqrt{1+\eta},0,0)$,
$\epsilon^\mu_y=\epsilon^{\prime\mu}_y=(0,0,1,0)$,
$\epsilon^\mu_z = \epsilon^{\prime \mu}_z(0,0,0,1)$
and $\epsilon^{\prime \mu}_x=(\sqrt{\eta},\sqrt{1+\eta},0,0)$.
For these
polarization four-vectors, the traces are given by:
\begin{eqnarray}
 Tr[ \ ]^{Bad}_{xx} & = & k^- \frac{\eta}{8} R
\ ,
\nonumber \\
Tr[ \ ]_{yy}^{Bad} & = & \  k^- (k^+ -p^+)^2 \ ,
\nonumber \\
Tr[ \ ]^{Bad}_{zz} & = & \frac18 ~k^- ~R \ ,
  \nonumber \\
Tr[ \ ]^{Bad}_{zx}&=& -  k^- \frac{ \sqrt{\eta} }{8} R
 \ ,
\label{traces} \end{eqnarray} where \begin{eqnarray} R= 4 ~Tr[
(\sla{k}-\sla{p^\prime} +m) \gamma^{+} (\sla{k}-\sla{p}+m)
\gamma^-] \ . \label{badtr}
\end{eqnarray}

The next step is to evaluate matrix elements of the
electromagnetic current for the "Bad" traces calculated above in
Eq. (\ref{badtr}). To perform the integration of the light-front
energy, $k^-$, in Eq. (\ref{jcurrent}) one needs to use   {\it the
pole dislocation method }, developed in
Refs.~\cite{Pacheco98,Pacheco992,Naus98,Pacheco99}, where it is
used  $q^+=\delta^+\rightarrow 0_+$. In detail, the pair terms or
Z-diagram contributions are given by: \bea J_{xx}^{+Z} =
 \lim_{\delta^+ \rightarrow 0}
\int d^4 k\frac{\theta(p^{\prime+}-k^+) \theta(k^+-p^{+})Tr[\
]^{Bad}_{xx}}{[1][2][3][4][5][6][7]} \ ;
\nonumber \\
J_{yy}^{+Z}  =
 \lim_{\delta^+ \rightarrow 0}
\int d^4 k\frac{\theta(p^{\prime+}-k^+) \theta(k^+-p^{+}) Tr[\
]^{Bad}_{yy}}{[1][2][3][4][5][6][7]} \ ;
\nonumber \\
J_{zx}^{+ Z}  =
 \lim_{\delta^+ \rightarrow 0}
\int d^4 k\frac{\theta(p^{\prime+}-k^+) \theta(k^+-p^{+}) Tr[\
]^{Bad}_{zx}]}{[1][2][3][4][5][6][7]} \ ;
\nonumber \\
J_{zz}^{+ Z } =
 \lim_{\delta^+ \rightarrow 0}
 \int d^4 k
\frac{ \theta(p^{\prime+}-k^+) \theta(k^+-p^{+})Tr[ \
]^{Bad}_{zz}}{[1][2][3][4][5][6][7]} \ . \label{currents} \eea The
results shown above correspond only to the computation of the
$k^-$ integration  for  $p^+ < k^+ < p^{\prime + }$ with $ p^{
\prime + }=p^+ + \delta^+$, where the pair term contribution to
the plus component of the electromagnetic current appears
~\cite{Pacheco98,Pacheco992}. The terms of the form  $k^{- (m+1)}
(p^+ -k^+)^n$ in Eq.~(\ref{currents}) are zero in the limit
$\delta\rightarrow  0_+$ if $m < n$~\cite{Pacheco992}. Therefore,
one immediately get that $J_{yy}^{+Z}=0$, which does not have a
pair term contribution, as has been already verified explicitly in
~\cite{Pacheco97,Pacheco992}.

The denominators in Eq.~(\ref{currents}), written with light-front
momentum are given by:
 \bea \relax [1] & = & k^+ (k^{-} - \frac{f_1-\imath
\epsilon}{k^+}) \ ;
\nonumber \\
\relax
[ 2 ]  & = &  (p^{+} -  k^+) (p^{-} - k^{-}
- \frac{f_2-\imath \epsilon}{p^+ - k^+})  \ ;  \nonumber \\
\relax
[ 3 ] & = &  (p^{\prime +} -  k^+) (p^{\prime -} - k^{-}
- \frac{f_3-\imath \epsilon}{p^{\prime +} - k^+}) \ ;  \nonumber \\
\relax
[ 4 ] & = &  (p^{+} -  k^+) (p^{-} - k^{-}
- \frac{f_4-\imath \epsilon}{p^{+} - k^+}) \ ; \nonumber \\
\relax
[ 5 ] & = &  (p^{\prime +} -  k^+) (p^{\prime -} - k^{-}
- \frac{f_5-\imath \epsilon}{p^{\prime +} - k^+}) \ ;  \nonumber \\
\relax
[ 6 ] & = &  (p^{+} -  k^+) (p^{-} - k^{-}
- \frac{f_6 - \imath \epsilon}{p^{\prime +} - k^+})  \ ; \nonumber \\
\relax [ 7 ]  & = &  (p^{\prime +} -  k^+) (p^{\prime -} - k^{-} -
\frac{f_7 - \imath \epsilon}{p^{\prime +} - k^+})  \  \ ; \relax
\eea where the functions $f_i$ are \bea
f_1 & = &  k_{\perp}^2+m^2  \ ; \nonumber  \\
f_2 & = & (k -p)_{\perp}^2 + m^2 \ ;  \nonumber  \\
f_3 & = & (k -p^{\prime})_{\perp}^2 + m^2 \ ;  \nonumber \\
f_4 & = & (k -p)_{\perp}^2 + m^2_{R}  \ ;  \nonumber \\
f_5 & = & (k -p^{\prime})_{\perp}^2 + m^2_{R} \ ;  \nonumber \\
f_6 & = & (k -p)_{\perp}^2 + m^2_{R}  \ ;  \nonumber \\
f_7 & = & (k -p^{\prime})_{\perp}^2 + m^2_{R}   \ . \eea

A final comment is appropriate here, the interval $0 < k^+ <p^+ $
in the integration of Eq.~(\ref{currents}) implies in a
non-vanishing result of  the $k^-$ integration, where the residue
receives contribution from the pole at $k^-=(f_1-\imath
\epsilon)/k^+$. This result is part of the valence contribution to
the electromagnetic current.

\section{Form factors and pair terms}

The angular condition~\cite{Pacheco97,Inna84,Inna89} satisfied by
the matrix elements (m.e.) of $J^+$ in the Breit-frame with
$q^+=0$, due to the requirements of rotational symmetry and parity
conservation is written as  \be \Delta(q^2)=(1+2 \eta)
I^{+}_{11}+I^{+}_{1-1} - \sqrt{8 \eta} I^{+}_{10} - I^{+}_{00} = 0
\  , \label{eq:ang} \ee with the m.e. of the plus component of the
current evaluated the light-cone polarization states, denoted as
$I^+_{m^\prime m}$. Different prescriptions to extract the form
factors choose three m.e. among the four independent ones (if only
the valence contribution is evaluated), or any other three
linearly independent combinations of matrix elements. The angular
condition in the instant form spin basis takes a particularly
simple form \be J^{+}_{yy}=J^{+}_{zz} \ . \ee The prescription
suggested by Grach and Kondratyuk~\cite{Inna84} eliminates the
matrix element $I^{+}_{00}$ from the computation of the form
factors using the angular condition.

The electromagnetic form factors are linear combinations of the
matrix elements of the electromagnetic current, and using the
prescription~\cite{Inna84}, one has \bea G_0^{GK} & = &
\frac{1}{3}[J_{xx}^{+}  + J_{yy}^{+} \ (2- \eta) + \eta
J_{zz}^{+}] \ ;
\nonumber \\
G_1^{GK} & = & [J_{yy} - J_{zz}^{+} - \frac{J_{zx}^+}{\eta}] \ ;
\nonumber \\
G_2^{GK}  & = & \frac{\sqrt{2}}{3}[J_{xx}^{+} + J_{yy}^{+} \ (-1-
\eta)+ \eta  J_{zz}^{+}] \ ; \label{ffactors} \eea where the
transformation of the light-cone to the instant form polarization
states were performed.

Now, due to the property of the traces of Eq.~(\ref{traces}),
i.e., \bea
Tr[ \ ]^{Bad}_{xx} & = &  -\eta \  Tr[ \ ]^{Bad}_{zz}  \ ;  \nonumber  \\
Tr[ \  ]^{Bad}_{zx} & = & -\sqrt{\eta} \ Tr[ \ ]^{Bad}_{zz}  \ ;
\label{vip} \eea one immediately derives from Eq.~(\ref{currents})
that \bea
J^{+Z}_{xx} & = &  -\eta \  J^{+Z}_{zz}  \ ;  \nonumber  \\
J^{+Z}_{zx} & = & -\sqrt{\eta} \ J^{+Z}_{zz}  \ . \label{vip1}
\eea Substituting these matrix elements in Eq.~(\ref{ffactors}),
we compute the contribution of the pair terms to the form factors:
\bea G_0^{GK,Z} & = & \frac{1}{3}[J_{xx}^{+Z} + \eta J_{zz}^{+Z}]=
0 \ ;
\nonumber \\
G_1^{GK,Z} & = & [-J_{zz}^{+Z} -\frac{J_{zx}^{+Z}}{\eta}] =0 \ ;
\nonumber    \\
G_2^{GK,Z}  & = & \frac{\sqrt{2}}{3}[J_{xx}^{+Z} + \eta
J_{zz}^{+Z}] = 0; \label{fiffactor} \eea where we have also used
that $J_{yy}^{+Z} = 0$.

The vanishing of the contribution of the pair terms as seen in
Eqs.~(\ref{fiffactor}), explains part of the results obtained by
the authors in Ref.~\cite{Pacheco97}, where it was found by
numerical analysis, that the prescription of Grach and Kondratyuk
with only the valence contribution reproduced the covariant
calculation.

\section{Conclusion}

In summary, we have analyzed the rotational symmetry properties of
the electromagnetic current in an analytical light-front model of
a spin-1 composite particle. The plus component of the current in
the Drell-Yan frame does not transform properly under rotations if
only the valence contribution is considered in the model. This
problem forbids a unique calculation of the electromagnetic
form-factors, due to the violation of the angular condition.

We have analyzed in deep, a particular prescription suggested by
Grach and Kondratyuk to extract the form factors from the plus
component of the current, and we have shown that it eliminates the
contributions from pair diagrams or zero modes, which if not
evaluated properly cause the violation of the rotational symmetry.
We have addressed this problem in an analytical and covariant
model of a spin-1 composite particle using a {\it a pole
dislocation method} to calculate the pair terms which survive the
limit of $q^+\rightarrow 0_+$. We have shown, in a direct and
compact form, the cancellation of the pair contribution in the
evaluation of the form factors using the prescription from
Ref.~\cite{Inna84}, which also happens using the instant form
polarization states in the cartesian representation, generalizing
previous findings ~\cite{JI2001}.  In a future work, we intend to
analyze with our methods the above prescription,  to extract form
factors from the electromagnetic current, with a more general form
of the coupling of the composite vector particle to the fermions.
The goal is to apply the results of such analysis to the study of
the $\rho$-meson or deuteron elastic photo-absorption processes.

\bigskip

\nl{\bf Acknowledgments}
\medskip

 This work was supported in part by the Brazilian
agencies FAPESP (Funda\c{c}\~ao de Amparo a Pesquisa do Estado de
S\~ao Paulo) and CNPq (Conselho Nacional de Desenvolvimento
Ci\^entifico e T\'ecnologico).

\end{multicols}


\begin{thebibliography}{99}

\bibitem{Terentev76} M.V. Terent\'ev,
Sov. J. Nucl. Phys. {\bf 24}, 106 (1976); L.A. Kondratyuk and M.V.
Terent\'ev, Sov. J. Nucl. Phys. {\bf 31}, 561 (1980).

\bibitem{Brodsky98}
S. J. Brodsky, H.-C. Pauli, and S. S. Pinsky, Phys. Rep.
{\bf 301}, 299 (1998).


\bibitem{Pacheco97}J.P.B.C. de Melo and T. Frederico,
Phy. Rev. C{\bf 55}, 2043 (1997).


\bibitem{Pacheco98}J.P.B.C. de Melo, J.H.0.Sales,
T. Frederico, and P.U. Sauer, Nucl. Phys. A{\bf 631 }, 574c (1998).

\bibitem{Pacheco992}J.P.B.C. de Melo, T. Frederico, H.W.L. Naus, and
P.U. Sauer, Nucl. Phys. A{\bf 660}, 219 (1999).

\bibitem{JI2001}B.L.G. Bakker and C.R. Ji,
Phy. Rev. D{\bf 65},  116001 (2002).

\bibitem{Naus98} H.W.L. Naus, J.P.B.C. de Melo, and
T. Frederico,  Few-Body Systems {\bf 24}, 99 (1998).

\bibitem{Pacheco99}
J.P.B.C. de Melo, H.W. L. Naus, and T. Frederico,
Phy. Rev. C{\bf 59}, 2278 (1999).

\bibitem{Inna84}I.L.Grach and  L.A.Kondratyuk,
Sov. J. Nucl. Phys. {\bf 38}, 198 (1984).

\bibitem{Inna89}I.L.Grach, L.A. Kondratyuk, and M.Strikman,
Phys. Rev. Lett. {\bf 62}, 387 (1989).

\bibitem{Pacheco2002}
J.P.B.C. de Melo, T. Frederico, E. Pace, and G. Salm\'e, Nucl.
Phys. A{\bf 707}, 399 (2002);  Braz. Jour. of Phys. {\bf 33}, 301
(2003).

\bibitem{Frankfurt79} L.L. Frankfurt and M. Strikman,
Nucl. Phys. B{\bf 148}, 107 (1979); Phys. Rep. {\bf 76}, 215
(1981).

\end{thebibliography}
\end{document}